\documentclass[showpacs,aps,prd,reprint,superscriptaddress,nofootinbib,longbibliography]{revtex4-1}
\usepackage[colorlinks=true, pdfstartview=FitV, linkcolor=magenta,citecolor=blue, urlcolor=magenta,
bookmarks=true, bookmarksnumbered=true, breaklinks]{hyperref}
\usepackage[dvipdfmx]{graphicx}
\usepackage{url}
\usepackage{amsmath,amssymb,bm,color,longtable,mathrsfs,amsfonts,slashed,ulem}

\newcommand{\Slash}[1]{{\ooalign{\hfil/\hfil\crcr$#1$}}}

\begin{document}
\begin{flushright}
\end{flushright}

\title{Decays of Roper-like singly heavy baryons in a chiral model}

\author{Daiki~Suenaga}
\email[]{suenaga@rcnp.osaka-u.ac.jp}
\affiliation{Research Center for Nuclear Physics,
Osaka University, Ibaraki, 567-0048, Japan }

\author{Atsushi~Hosaka}
\email[]{{hosaka@rcnp.osaka-u.ac.jp}}
\affiliation{Research Center for Nuclear Physics,
Osaka University, Ibaraki, 567-0048, Japan }
\affiliation{Advanced Science Research Center, Japan Atomic Energy Agency (JAEA), Tokai 319-1195, Japan}

\date{\today}

\begin{abstract}
We present a chiral model describing decays of the Roper-like $\Lambda_c(2765)$ and $\Xi_c(2970)$ as well as those of the heavy quark spin-doublet $\{\Sigma_c(2455),\Sigma_c(2520)\}$ and $\{\Xi_c',\Xi_c(2645)\}$, by focusing on chiral representations of diquarks inside the heavy baryons. 
Based on our chiral model together with the heavy quark spin symmetry, we derive a relation satisfied by axial charges of the heavy baryons that controls magnitude of their one pion (kaon) decays. Besides, it is shown that the large decay widths of the Roper-like $\Lambda_c(2765)$ and $\Xi_c(2970)$, which cannot be implemented by conventional non-relativistic quark models, are explained by reasonable values of the axial charges. In addition, we apply the chiral model to bottom baryons and predict decay widths of the undiscovered Roper-like $\Xi_b$. We expect that our present investigation leads to elucidation of dynamical properties of the excited heavy baryons from the viewpoint of chiral symmetry.
\end{abstract}


\maketitle

\section{Introduction}
\label{sec:Introduction}

With the help of recent development of hadron experiments at, e.g., KEK, LHC and SLAC, properties of singly heavy baryons have been unveiled. Since singly heavy baryons include only one heavy quark ($c$ or $b$ quark) whose mass is larger than the typical energy scale $\Lambda_{\rm QCD}$ of Quantum Chromodynamics (QCD), those baryons are useful to understand properties of the remaining diquark inside the baryons which is colorful~\cite{manohar2000heavy}.

Among singly heavy baryons, the Roper-like baryons such as $\Lambda_c(2765)$ and $\Xi_c(2970)$ for the charm sector are particularly of interest. 
Those baryons are heavy quark spin-singlet (HQS-singlet) whose spin and parity are $J^P=1/2^+$~\cite{Belle:2020tom},\footnote{Although the spin and parity of $\Lambda_c(2765)$ have not been established, they seem to be $J^P=1/2^+$~\cite{Arifi:2020ezz}.} which are the same as the corresponding ground-state $\Lambda_c$ and $\Xi_c$. Their masses are larger than the ground-state ones by about $500$ MeV~\cite{Zyla:2020zbs}. This mass difference is similar to the Roper resonance $N(1440)$ in the nucleon sector~\cite{Roper:1964zza}, and such a universality of the mass difference seems to persist to the bottom sector~\cite{CMS:2020zzv,Aaij:2020rkw}. In addition, it is known that the large decay widths of $\Lambda_c(2765)$ and $\Xi_c(2970)$ whose values are tens of MeV are not easily explained by the conventional non-relativistic quark model~\cite{Copley:1979wj,Yoshida:2015tia} giving the widths of only a few MeV~\cite{Nagahiro:2016nsx}.\footnote{In Ref.~\cite{Arifi:2021orx}, we showed that a quark model including relativistic corrections can explain the large decay widths of Roper-like baryons. In addition, the $^3P_0$ model can reproduce the decay widths as well as the masses of them satisfactorily~\cite{Chen:2016iyi}. } 

In this paper, we study decay properties of the heavy baryons, especially focusing on the large decay widths of the Roper-like baryons, by means of a chiral model together with the heavy quark spin symmetry~\cite{Kawakami:2018olq,Kawakami:2019hpp,Harada:2019udr,Dmitrasinovic:2020wye,Kawakami:2020sxd,Suenaga:2021qri}. The chiral model describes the heavy baryons based on chiral symmetry, which is one of the fundamental symmetries in QCD, exhibited by the diquark inside the baryons. Hence, the model is useful to examine interaction properties among the heavy baryons and light pseudo-scalar mesons, regarded as the Nambu-Goldstone (NG) bosons associated with the chiral symmetry breaking. In other words, our present investigation leads to the better understanding of such dynamical natures of the heavy baryons from the viewpoint of chiral symmetry~\cite{Kawakami:2018olq,Kawakami:2019hpp,Kawakami:2020sxd}.

Focusing on the Roper-like $\Lambda_c(2765)$, this baryon decays into the ground-state $\Lambda_c$ by emitting two pions. The experimental observation suggests that the direct two-pion emission decay is not important~\cite{Belle:2006xni}, and hence, main decay modes are the sequential processes via the heavy quark spin-doublet (HQS-doublet) baryons such as $\{\Sigma_c(2455), \Sigma_c(2520)\}$: $\Lambda_c(2765) \to \Sigma_c(2455)\pi \to \Lambda_c\pi\pi$ and $\Lambda_c(2765) \to \Sigma_c(2455)\pi \to \Lambda_c\pi\pi$. We assume that such a selection rule applies to $\Xi_c(2970)$ as well. For this reason, here, we do not take into account the direct decays of the Roper-like $\Lambda_c(2765)$ and $\Xi_c(2970)$. As for the sequential processes, we evaluate the decay widths by computing only the primary processes approximately by ignoring the secondary ones as done in Refs.~\cite{Nagahiro:2016nsx,Arifi:2021orx,Chen:2016iyi}.

In Ref.~\cite{Suenaga:2021qri}, based on a chiral model we argued that the Roper-like $\Lambda_c(2765)$, $\Xi_c(2970)$ and the ground-state $\Lambda_c$, $\Xi_c$ are superpositions of a three-quark state $Qqq$ and a pentaquark states $Qqq\bar{q}q$, and the former was shown to be mostly $Qqq\bar{q}q$ while the latter $Qqq$.\footnote{In Ref.~\cite{Suenaga:2021qri} we showed that our chiral model can naturally explain the suppression of direct decays of the Roper-like baryons.} In this paper, we generalize the discussions in Ref.~\cite{Suenaga:2021qri} by focusing on chiral representations of the diquark. Besides, we extend the model by incorporating HQS-doublet baryons such as $\{\Sigma_c(2455), \Sigma_c(2520)\}$ and $\{\Xi_c',\Xi_c(2645)\}$ carrying $J^P=1/2^+,3/2^+$. Based on such a model, we examine decays of the Roper-like $\Lambda_c(2765)$, $\Xi_c(2970)$ as well as those of the HQS-doublet $\{\Sigma_c(2455), \Sigma_c(2520)\}$ and $\{\Xi_c',\Xi_c(2645)\}$.

With respect to the heavy quark spin symmetry, investigation of the bottom baryons are more useful. For the bottom baryons, only the Roper-like $\Lambda_b$ was discovered and the Roper-like $\Xi_b$ is still missing~\cite{Zyla:2020zbs}. Thus, in this paper we apply our chiral model to the bottom baryons as well, and present predictions of decay widths of the undiscovered Roper-like $\Xi_b$.

Our present paper is organized as follow. In Sec.~\ref{sec:Model} we construct our chiral Lagrangian to investigate decays of the heavy baryons, indicating important points presented in Ref.~\cite{Suenaga:2021qri}. Based on the Lagrangian, in Sec.~\ref{sec:Analysis} we show analytic expressions of the decay widths. In Sec.~\ref{sec:Results} numerical results of the decay widths for the charmed baryons are shown. Besides, in Sec.~\ref{sec:Bottom} we apply the model to bottom baryons. In Sec.~\ref{sec:Discussion}, we comment on the important relation derived in our chiral model that axial charges of the heavy baryons satisfy, and discuss decays of the parity partners (or the so-called chiral partners)~\cite{Harada:2019udr,Kawakami:2020sxd}. And in Sec.~\ref{sec:Conclusions} we summarize the present study.

\section{Model}
\label{sec:Model}

The Roper-like baryons $\Lambda_c(2765)$ and $\Xi_c(2970)$ decay into the HQS-doublet baryons $\{\Sigma_c(2455),\Sigma_c(2520)\}$ and $\{\Xi_c',\Xi_c(2645)\}$ by emitting a pion (and a kaon). In this section we construct a Lagrangian describing those decays based on chiral symmetry and heavy quark spin symmetry.

\subsection{HQS-singlet baryons}
\label{sec:HQSSinglet}

In this subsection, we introduce HQS-singlet baryon fields describing both the Roper-like $\Lambda_c(2765)$ and $\Xi_c(2970)$, and the ground-state $\Lambda_c$ and $\Xi_c$. For this purpose, to begin with, we focus on the diquarks inside the baryons.

The diquarks inside the HQS-singlet baryons are Lorentz scalar. Thus, possible interpolating fields for the diquarks including only two quarks are~\cite{Liu:2011xc,Harada:2019udr}
\begin{eqnarray}
(d_R)^a_i &\sim& \epsilon_{ijk}\epsilon^{abc}(q_R^T)^b_jC(q_R)^c_k\ ,  \nonumber\\
(d_L)^a_i &\sim& \epsilon_{ijk}\epsilon^{abc}(q_L^T)^b_jC(q_L)^c_k \ , \label{Diquark1}
\end{eqnarray}
where $q_R=\frac{1+\gamma_5}{2}q$ and $q_L=\frac{1-\gamma_5}{2}q$ are right-handed and left-handed quark fields, respectively. The symbol $C$ is the charge-conjugation operator $C=i\gamma^2\gamma^0$, 
and the superscript $a,b,\cdots$ and subscript $i,j,\cdots$ represent color and flavor indices, respectively. Hence, the chiral representations of the diquarks in Eq.~(\ref{Diquark1}) are
\begin{eqnarray}
d_R &\sim& ({\bm 1},\bar{\bm 3}) \ , \ \ d_L\sim(\bar{\bm 3},{\bm 1})\ . \label{ChiralDiquark1}
\end{eqnarray}

In order to describe the Roper-like baryons $\Lambda_c(2765)$ and $\Xi_c(2970)$ in addition to the ground-state ones $\Lambda_c$ and $\Xi_c$ collectively, we need to introduce the other diquarks $d_R'$ and $d_L'$, whose Lorentz structures, and accordingly chiral representations, are identical to those of $d_R$ and $d_L$. Here, we assign the chiral representations of $d'_R$ and $d_L'$ as
\begin{eqnarray}
d'_R&\sim& (\bar{\bm 3},{\bm 1}) \ , \ \ d'_L\sim({\bm 1},\bar{\bm 3})\ , \label{ChiralDiquark2}
\end{eqnarray}
where the representations are flipped from those of $d_R$ and $d_L$ in Eq.~(\ref{ChiralDiquark1}), following Ref.~\cite{Suenaga:2021qri}.\footnote{In this reference, we introduced the so-called {\it mirror diquarks}
\begin{eqnarray}
(d'_R)_i^a &\sim&\epsilon_{jkl}\epsilon^{abc}(q_R^T)^b_kC(q_R)^c_l [(\bar{q}_L)^d_i(q_R)^d_j] \ , \nonumber\\
(d'_L)_i^a &\sim&  \epsilon_{jkl}\epsilon^{abc}(q_L^T)^b_kC(q_L)^c_l[(\bar{q}_R)^d_i(q_L)^d_j] \ , \label{MirrorDiquark}
\end{eqnarray}
which are regarded as tetraquark states for $d_R'$ and $d_L'$. The mirror diquarks give the flipped chiral representations as in Eq.~(\ref{ChiralDiquark2}), because of the $\bar{q}_Lq_R$ ($\bar{q}_Rq_L$) component in Eq.~(\ref{MirrorDiquark}).} 

Interpolating fields of HQS-singlet baryons composed of $d_R$, $d_L$, $d_R'$ and $d_L'$ are given by attaching a heavy quark $Q$ to the diquarks as
\begin{eqnarray}
(B_R)_i \sim Q^a (d_R)_i^a \ \ &,&\ \ (B_L)_i \sim Q^a(d_L)_i^a \ , \nonumber\\
(B_R')_i \sim Q^a (d_R')_i^a \ \  &,& \ \ (B_L')_i \sim Q^a(d_L')_i^a \ . \label{BRBLFields}
\end{eqnarray}
From the baryon fields in Eq.~(\ref{BRBLFields}), we can construct a chiral invariant Lagrangian with Eqs.~(\ref{ChiralDiquark1}) and~(\ref{ChiralDiquark2}). Here, for later use we define anti-symmetric matrices of the baryons in terms of the flavor space as~\cite{Kawakami:2019hpp}
\begin{eqnarray}
({\cal B}_R)_{ij} \equiv  \epsilon_{ijk}(B_R)_k \ \ &,& \ \ ({\cal B}_L)_{ij} \equiv  \epsilon_{ijk}(B_L)_k \ , \nonumber\\
({\cal B}_R')_{ij} \equiv  \epsilon_{ijk}(B_R')_k \ \ &,& \ \ ({\cal B}_L')_{ij} \equiv  \epsilon_{ijk}(B_L')_k \ . \label{BRBLMatrix}
\end{eqnarray}
The anti-symmetric structures in Eq.~(\ref{BRBLMatrix}) manifestly reflect the anti-symmetric chiral representations in Eqs.~(\ref{ChiralDiquark1}) and~(\ref{ChiralDiquark2}). Hence, the $SU(3)_L\times SU(3)_R$ chiral transformation laws of the baryons defined in Eq.~(\ref{BRBLMatrix}) become simply
\begin{eqnarray}
{\cal B}_R \overset{\rm ch.}{\to} g_R{\cal B}_R g_R^T \ \ &,&\ \ {\cal B}_L  \overset{\rm ch.}{\to}  g_L{\cal B}_L g_L^T \ ,\nonumber\\
{\cal B}_R'  \overset{\rm ch.}{\to}  g_L {\cal B}_R' g_L^T \ \ &,&\ \  {\cal B}_L'  \overset{\rm ch.}{\to}  g_R{\cal B}_L' g_R^T \ , \label{BRBLChiral}
\end{eqnarray}
without changing the representations in Eqs.~(\ref{ChiralDiquark1}) and~(\ref{ChiralDiquark2}), where $g_{R(L)}\in SU(3)_{R(L)}$. Besides, the parity transformation laws read
\begin{eqnarray}
{\cal B}_R  \overset{\rm P}{\to}   -\gamma^0{\cal B}_L \ \ &,&\ \ {\cal B}_L  \overset{\rm P}{\to} -\gamma^0{\cal B}_R \ ,\nonumber\\
{\cal B}_R'  \overset{\rm P}{\to} -\gamma^0{\cal B}_L' \ \ &,&\ \  {\cal B}_L'  \overset{\rm P}{\to} -\gamma^0{\cal B}_R'  \ . \label{BRBLParity}
\end{eqnarray}

The basis of ${\cal B}^{(\prime)}_{R(L)}$ is useful to construct a chiral invariant Lagrangian, since the chiral representations are manifestly exhibited. They are, however, not parity eigenstates as seen from Eq.~(\ref{BRBLParity}). The parity eigenstates are given by
\begin{eqnarray}
{\cal B}^{(\prime)}_+ = \frac{1}{\sqrt{2}}({\cal B}^{(\prime)}_R-{\cal B}^{(\prime)}_L) \ , \ \ {\cal B}^{(\prime)}_- = \frac{1}{\sqrt{2}}({\cal B}^{(\prime)}_R+{\cal B}^{(\prime)}_L) \ , \label{BParity}
\end{eqnarray}
where the subscript ``$+$'' or ``$-$'' corresponds to the parity eigenvalue. Thus, defining parity eigenstates $B_\pm^{(\prime)}$ from $B_{R(L)}^{(\prime)}$ in Eq.~(\ref{BRBLFields}) in a similar manner to Eq.~(\ref{BParity}), the matrix ${\cal B}^{(\prime)}_\pm$ is explicitly expressed as
\begin{eqnarray}
{\cal B}_{\pm} &=& \left(
\begin{array}{ccc}
0 & B_{\pm,i=3} & -B_{\pm, i=2} \\
-B_{\pm,i=3} & 0 & B_{\pm,i=1} \\
B_{\pm,i=2} & -B_{\pm,i=1} & 0 \\
\end{array}
\right) \ ,\nonumber\\
{\cal B}'_{\pm} &=& \left(
\begin{array}{ccc}
0 & B'_{\pm,i=3} & -B'_{\pm,i=2} \\
-B'_{\pm,i=3} & 0 & B'_{\pm,i=1} \\
B'_{\pm,i=2} & -B'_{\pm,i=1} & 0 \\
\end{array}
\right) \ .
\end{eqnarray}

In Ref.~\cite{Suenaga:2021qri}, based on the spontaneous breakdown of chiral symmetry, it was shown that in general the mass eigenstates are given by superpositions of ${B}_\pm$ and ${B}'_\pm$. That is, the mass eigenstates read
\begin{eqnarray}
\left(
\begin{array}{c}
B_{\pm,i}^L \\
B_{\pm,i}^H \\
\end{array}
\right) = \left(
\begin{array}{cc}
\cos\theta_{B_{\pm}^i} & \sin\theta_{B_{\pm}^i} \\
-\sin\theta_{B_{\pm}^i} & \cos\theta_{B_{\pm}^i} \\
\end{array}
\right)\left(
\begin{array}{c}
B_{\pm,i} \\
B_{\pm,i}' \\
\end{array}
\right)\ , \label{Mixing}
\end{eqnarray}
where the superscripts $L$ and $H$ refer to lower and higher masses. In later analysis, we will only focus on the positive-parity $B_+^L$ and $B_+^H$, where $B_+^L$ corresponds to the ground-state $\Lambda_c$ and $\Xi_c$ while $B_+^H$ the Roper-like $\Lambda_c(2765)$ and $\Xi_c(2970)$. Explicitly, $B_+^L$ and $B_+^H$ correspond to the experimentally observed baryons as
\begin{eqnarray}
&&B^L_{+,i=1} = \Xi_c^0\ \ ,\ \ B^H_{+,i=1} = \Xi_c(2970)^0\ , \nonumber\\
&& B^L_{+,i=2},  =\Xi_c^+\ \ , \ \ B^H_{+,i=2} = \Xi_c(2970)^+\ , \nonumber\\
&& B^L_{+,i=3} = \Lambda_c^+ \  \ , \ \  B^H_{+,i=3}= \Lambda_c(2765)^+\ . \label{BAssign}
\end{eqnarray}

\subsection{HQS-doublet baryons}
\label{sec:HQSDoublet}

In this subsection, we introduce HQS-doublet baryon fields describing $\{\Sigma_c(2455), \Sigma_c(2520)\}$ and $\{\Xi_c',\Xi_c(2645)\}$. The diquark inside the baryons must carry spin one since those baryons belong to the spin-doublet with the total spin $J=1/2, 3/2$. The simplest diquark field constructed by two quarks is Lorentz vector, which is uniquely given by~\cite{Liu:2011xc,Kawakami:2019hpp,Harada:2019udr}
\begin{eqnarray}
d_{(i,j)}^{a,\mu} \sim \epsilon^{abc}(q_L^T)^b_j C\gamma^\mu (q_R)^c_i \ . \label{Diquark3}
\end{eqnarray}
Here, it should be noted that the Lorentz vector diquarks including two $q_L$'s ($q_R$'s) do not exist due to the Pauli principle.

Similarly to Eq.~(\ref{BRBLFields}), interpolating fields of HQS-doublet baryons $S^\mu$ are given by attaching a heavy quark $Q$ to the diquark in Eq.~(\ref{Diquark3}), i.e.,
\begin{eqnarray}
S^\mu_{ij} \sim Q^ad_{(i,j)}^{a,\mu}\ .
\end{eqnarray}
Hence, the $SU(3)_L\times SU(3)_R$ chiral and parity transformation laws of the baryon $S^\mu$ read
\begin{eqnarray}
S^\mu \overset{\rm ch.}{\to} g_R S^\mu g_L^T \  \ \ , \ \ \ S^\mu \overset{\rm P}{\to}-\gamma^0 S^T_\mu \ , \label{STrans}
\end{eqnarray}
respectively, from the diquark~(\ref{Diquark3}). 
When we separate $S^\mu$ into two matrices which are flavor symmetric and anti-symmetric, i.e., the flavor sextet $S^\mu_6$ and anti-triplet $S_{\bar{3}}^\mu$ parts as
\begin{eqnarray}
S^\mu = S_6^\mu + S_{\bar{3}}^\mu \ ,
\end{eqnarray}
the parity transformation in Eq.~(\ref{STrans}) yields
\begin{eqnarray}
S_6^\mu  \overset{\rm P}{\to} -\gamma^0 S_{6\mu} \ \ \ , \ \ \ S_{\bar{3}}^\mu  \overset{\rm P}{\to} \gamma^0 S_{\bar{3}\mu} \ . \label{S6S3Parity}
\end{eqnarray}
In other words, $S^\mu_6$ contains positive-parity baryons while $S^\mu_{\bar{3}}$ negative-party ones.\footnote{Our definition of the diquark~(\ref{Diquark3}) reads $d^{a\mu}\overset{\rm P}{\to}-(d^a_\mu)^T$ under the proper parity transformation having an extra minus sign.} Thus, we can assign the HQS-doublet baryons as
\begin{eqnarray}
S_6^\mu = \left(
 \begin{array}{ccc}
\Sigma_c^{I_z=1} & \frac{1}{\sqrt{2}}\Sigma_c^{I=0} & \frac{1}{\sqrt{2}}\Xi_c'^{I=1/2} \\
\frac{1}{\sqrt{2}}\Sigma_c^{I=0} & \Sigma_c^{I=-1} & \frac{1}{\sqrt{2}}\Xi_c'^{I=-1/2} \\
\frac{1}{\sqrt{2}}\Xi_c'^{I=1/2} & \frac{1}{\sqrt{2}}\Xi_c'^{I=-1/2} & \Omega_c \\
\end{array}
\right)^\mu\ , \nonumber\\ \label{S6Matrix}
\end{eqnarray}
and
\begin{eqnarray}
 S_{\bar{3}}^{\mu} = \frac{1}{\sqrt{2}}\left(
 \begin{array}{ccc}
0& \Lambda_{c1} & \Xi_{c1}^{I=1/2} \\
- \Lambda_{c1} & 0 & \Xi_{c1}^{I=-1/2} \\
-\Xi_{c1}^{I=1/2} & -\Xi_{c1}^{I=-1/2}&0\\
\end{array}
\right)^\mu\ . \nonumber\\ \label{S3BarMatrix}
\end{eqnarray}
For later use, following the useful field definition proposed in Ref.~\cite{Falk:1991nq}, we decompose $S_{6(\bar{3})}^\mu$ into $J=3/2$ part $S_{6(\bar{3})}^{*\mu}$ and $J=1/2$ part $S_{6(\bar{3})}$ as
\begin{eqnarray}
S_{6(\bar{3})}^\mu = S_{6(\bar{3})}^{*\mu} + \frac{1}{\sqrt{3}}(\gamma^\mu + v^\mu)\gamma_5 S_{6(\bar{3})}\ , \label{SpinDec}
\end{eqnarray}
with $v^\mu$ being a velocity of the baryon.\footnote{The decomposition in Eq.~(\ref{SpinDec}) is done such that $ \gamma_\mu S^{*\mu}_{6(\bar{3})}=0$ as well as $ v_\mu S^{*\mu}_{6(\bar{3})}=0$ and $\Slash{v}S^{*\mu}_{6(\bar{3})}=S^{*\mu}_{6(\bar{3})}$ is always satisfied. In our present analysis we do not include off-shell contributions of $S^{*\mu}_{6(\bar{3})}$ to evaluate the decay widths, and $S^{*\mu}_{6(\bar{3})}$ is always on-shell. Thus, the off-shell parameter $z_i$~\cite{Krebs:2009bf} and the arbitrary parameter $A$~\cite{Moldauer:1956zz} do not change the following arguments.}

In later analysis, we will leave only the flavor sextet part $S^\mu_6$ where $\{\Sigma_c(2455),\Sigma_c(2520)\}$ and $\{\Xi_c',\Xi_c(2645)\}$ are contained.  
In the decomposition in Eq.~(\ref{SpinDec}), $\Sigma_c(2455)$ and $\Xi_c'$ belong to $S_6$ while $\Sigma_c(2520)$ and $\Xi_c(2645)$ belong to $S_6^{*\mu}$, and their assignment in the flavor space is understood from Eq.~(\ref{S6Matrix}).

\subsection{Interaction Lagrangian}
\label{sec:Lagrangian}

Here, we construct a Lagrangian describing interactions among the HQS-singlet baryons ${\cal B}^{(\prime)}_{R(L)}$ and the HQS-doublet ones $S^\mu$.

From the transformation laws in Eqs.~(\ref{BRBLChiral}),~(\ref{BRBLParity}) and~(\ref{STrans}) [or Eq.~(\ref{S6S3Parity})], an interaction Lagrangian which is invariant under the $SU(3)_L\times SU(3)_R$ chiral, $SU(2)_h$ heavy quark spin, and parity transformations reads
\begin{eqnarray}
{\cal L}_{\rm int} &=&  \frac{\sqrt{2}g_{A}}{f_0}{\rm Tr}\Big[\bar{\cal B}_R\partial_\mu \Sigma^\dagger S^{T\mu}+\bar{\cal B}_L\partial_\mu\Sigma S^\mu +{\rm h.c.}\Big] \nonumber\\
&+& \frac{\sqrt{2}g_{A}'}{f_0}{\rm Tr}\Big[\bar{\cal B}_L'\partial_\mu \Sigma^\dagger S^{T\mu}+\bar{\cal B}_R'\partial_\mu\Sigma S^\mu + {\rm h.c,}\Big]\ .\nonumber\\ 
 \label{LInt}
\end{eqnarray}
In this Lagrangian, 
\begin{eqnarray}
\Sigma=S+iP
\end{eqnarray}
is a light meson matrix containing a scalar nonet $S$ and a pseudo-scalar nonet $P$, and we have left interaction terms having one $\partial_\mu\Sigma$ or $\partial_\mu\Sigma^\dagger$. Besides, $f_0$ is an averaged pseudo-scalar decay constant: $f_0=\frac{f_\pi+f_K}{2}=101\, {\rm MeV}$, with $f_\pi=92.1\, {\rm MeV}$ and $f_K=110\, {\rm MeV}$~\cite{Zyla:2020zbs}.\footnote{Here, we have taken into account only $f_\pi$ and $f_K$ to estimate $f_0$ since the pseudo-scalar mesons relevant to decays of the Roper-like baryons are pions and kaons. When we estimate $f_0$ from $f_0=\frac{f_\pi+f_K+f_\eta}{3}$, the resultant value reads $f_0=107\, {\rm MeV}$ with $f_\eta=1.3f_\pi$~\cite{Gasser:1984gg}.} In this way $g_{A}$ and $g_{A}'$ become dimensionless. Making use of Eq.~(\ref{BParity}), the Lagrangian~(\ref{LInt}) turns into
\begin{eqnarray}
{\cal L}_{\rm int}  &=& \frac{2g_{A}}{f_0}{\rm Tr}\Big[-i\bar{\cal B}_+\partial_\mu PS_6^{\mu}-\bar{\cal B}_+\partial_\mu SS_{\bar{3}}^\mu + i \bar{\cal B}_-\partial_\mu PS_{\bar{3}}^{\mu}\nonumber\\
&+&  \bar{\cal B}_-\partial_\mu SS_6^{\mu} + {\rm h.c.}
\Big] +\frac{2g_{A}'}{f_0}{\rm Tr}\Big[-i\bar{\cal B}_+'\partial_\mu PS_6^{\mu} \nonumber\\
&-&\bar{\cal B}_+'\partial_\mu SS_{\bar{3}}^{\mu} + i\bar{\cal B}_-'\partial_\mu PS_{\bar{3}}^{\mu} + \bar{\cal B}_-'\partial_\mu S S_6^{\mu}+ {\rm h.c.} \Big]\ , \nonumber\\ \label{LIntParity}
\end{eqnarray}
In this Lagrangian, since $S_6^\mu$ and $S_{\bar{3}}^\mu$ carry the positive- and negative-parity HQS-doublets, respectively, the parity invariance of the Lagrangian derives a selection rule for their couplings; for instance, the coupling between $S_6^\mu$ and ${\cal B}_+^{(\prime)}$ is mediated by a pseudo-scalar meson $P$, while that between $S_{\bar{3}}^\mu$ and ${\cal B}_+^{(\prime)}$ is by a scalar meson $S$. Moreover, chiral symmetry demands that the couplings for parity partners~\cite{Harada:2019udr,Kawakami:2020sxd} of the heavy baryons should be commonly controlled by $g_A$ and $g_A'$. In fact, the couplings for ${\rm Tr}[\bar{S}_6^\mu\partial_\mu P{\cal B}_+]$ and ${\rm Tr}[\bar{S}_{\bar{3}}^\mu\partial_\mu P{\cal B}_-]$ are identical. Discussions related to this structure will be given in Sec.~\ref{sec:ParityPartner}.

In what follows, 
we focus on only the positive-parity heavy baryons in our chiral model~(\ref{LIntParity}). As mentioned in Sec.~\ref{sec:HQSSinglet} and Sec.~\ref{sec:HQSDoublet}, those baryons are provided by ${\cal B}_+^{(\prime)}$ and $S_6^\mu$. Hence, we can neglect ${\cal B}_-^{(\prime)}$ and $S_{\bar{3}}$ in Eq.~(\ref{LIntParity}), i.e., the Lagrangian is reduced to
\begin{eqnarray}
{\cal L}_{\rm int}  &=& i\frac{2g_{A}}{f_0}{\rm Tr}\Big[\bar{S}_6^\mu \partial_\mu P{\cal B}_+-\bar{\cal B}_+\partial_\mu PS_6^{\mu}\Big]\nonumber\\
&+& i \frac{2g_{A}'}{f_0}{\rm Tr}\Big[\bar{S}_6^\mu \partial_\mu P{\cal B}'_+-\bar{\cal B}'_+\partial_\mu PS_6^{\mu} \Big]\ , \label{LIntParity2}
\end{eqnarray}
where the pseudo-scalar nonet regarded as the NG bosons associated with the chiral symmetry breaking is explicitly given by
\begin{eqnarray}
P  &=& \frac{1}{\sqrt{2}}\left(
\begin{array}{ccc}
\frac{\eta_N+\pi^0}{\sqrt{2}} & \pi^+& K^+ \\
\pi^- & \frac{\eta_N-\pi^0}{\sqrt{2}}  & K^0 \\
K^-& \bar{K}^0 & \eta_S \\
\end{array} 
\right) \ .
\end{eqnarray}

Before closing this section, we mention axial charges of the heavy baryons. Couplings among the heavy baryons and an axial gauge field ${\cal A}_\mu$ can be read off by replacing the derivative $\partial_\mu \Sigma$ to the following covariant derivative, in the Lagrangian~(\ref{LInt}):
\begin{eqnarray}
D_\mu \Sigma = \partial_\mu\Sigma +\frac{i}{2}\{{\cal A}_\mu,\Sigma\} \ .\label{CovariantSigma}
\end{eqnarray}
That is, in terms of $S_6^\mu$ and $B_+^{(\prime)}$ we can read the couplings as
\begin{eqnarray}
{\cal L}_{\rm axial} &=& ig_{A}{\rm Tr}\Big[\bar{S}_6^\mu {\cal A}_\mu{\cal B}_+-\bar{\cal B}_+{\cal A}_\mu S_6^{\mu}\Big] \nonumber\\
&+& i g_{A}'{\rm Tr}\Big[\bar{S}_6^\mu {\cal A}_\mu {\cal B}'_+-\bar{\cal B}'_+{\cal A}_\mu S_6^{\mu} \Big]\ . \label{LAxial}
\end{eqnarray}
In obtaining Eq.~(\ref{LAxial}), we have assumed that the vacuum expectation value (VEV) of $\Sigma$ is diagonal in the flavor space as $\langle\Sigma\rangle = \frac{f_0}{2}{\bm 1}$. That is, we assume the $SU(3)_{L+R}$ symmetry is satisfied in our present investigation, and indeed this assumption seems to be reasonable as will be seen later. Equation~(\ref{LAxial}) implies that $g_{A}$ and $g_{A}'$ are understood to be the axial charges defined in this way.

\section{Analytic evaluation of decay widths}
\label{sec:Analysis}

Our chiral model~(\ref{LIntParity2}) describes the ground-state $\Lambda_c$ and $\Xi_c$, the Roper-like $\Lambda_c(2765)$ and $\Xi_c(2970)$, and the HQS-doublet $\{\Sigma_c(2450),\Sigma_c(2520)\}$ and $\{\Xi_c',\Xi_c(2645)\}$, together with the NG bosons in a simple form. Therefore, in Sec.~\ref{sec:Results} we will study decay widths of both the Roper-like and the HQS-doublet baryons, in order to fully examine dynamical properties of the baryons interacting with the NG bosons from chiral symmetry. Before presenting the numerical results, in this section we clarify our procedure by showing analytic expressions of decay widths of $\Lambda_c(2765) \to \Sigma_c(2455)\pi$, $\Lambda_c(2765) \to \Sigma_c(2520)\pi$,  and $\Sigma_c(2455)\to\Lambda_c\pi$, $\Sigma_c(2520)\to \Lambda_c\pi$ in detail.

First, we show how to compute decays of the Roper-like $\Lambda_c(2765)$. From the Lagrangian~(\ref{LIntParity2}), couplings describing interactions of $\Lambda_c(2765)$-$\Sigma_c(2455)$-$\pi$ and $\Lambda_c(2765)$-$\Sigma_c(2520)$-$\pi$ are read off as
\begin{eqnarray}
{\cal L}_{\rm int} &\to& 
\frac{i\sqrt{2}}{f_0}g_{A,3}^{+H} \Big[\big(\bar{\Sigma}_Q^{I=1}\big)^\mu  \partial_\mu\pi^+  {B}_{+,3}^{H}\nonumber\\
&-& \big(\bar{\Sigma}_Q^{I=0}\big)^\mu  \partial_\mu\pi^0  {B}_{+,3}^{H}  -   \big(\bar{\Sigma}_Q^{I=-1}\big)^\mu \partial_\mu\pi^-  {B}_{+,3}^{H} \Big]\nonumber\\
&+& \frac{i\sqrt{2}}{f_0}g_{A,3}^{+L} \Big[ \bar{B}_{+,3}^{L}\partial_\mu\pi^-(\Sigma_Q^{I=1})^\mu \nonumber\\
&-& \bar{B}_{+,3}^{L}\partial_\mu\pi^0(\Sigma_Q^{I=0})^\mu - \bar{B}_{+,3}^{L}\partial_\mu\pi^+(\Sigma_Q^{I=-1})^\mu\Big]  \  , \nonumber\\
\label{ModelReduced}
\end{eqnarray}
where we have used Eq.~(\ref{Mixing}) and have defined 
\begin{eqnarray}
g_{A,3}^{+H} &\equiv& g_{A} \sin\theta_{B_+^3}-g_{A}'\cos\theta_{B_+^3} \ , \nonumber\\
g_{A,3}^{+L} &\equiv& g_{A} \cos\theta_{B_+^3} + g_{A}'\sin\theta_{B_+^3} \ .\label{Couplings}
\end{eqnarray}
It should be noted that, in our notation, the Roper-like $\Lambda_c(2765)$ and the ground-state $\Lambda_c$ are denoted by $B_{+,3}^{H}$ and $B_{+,3}^{L}$, respectively, from Eq.~(\ref{BAssign}), while the HQS-doublet $\{\Sigma_c(2455),\Sigma_c(2520)\}$ is provided by $\big({\Sigma}_Q^{I=\pm1,0}\big)^\mu$. Hence, $g_{A,3}^{+H}$ and $g_{A,3}^{+L}$ in Eq.~(\ref{Couplings}) correspond to the axial charges related to $\Lambda_c(2765)$-$\{\Sigma_c(2455),\Sigma_c(2520)\}$ sector and $\Lambda_c$-$\{\Sigma_c(2455),\Sigma_c(2520)\}$ sector, respectively. 

\begin{figure}[tb]
\centering
\includegraphics*[scale=0.53]{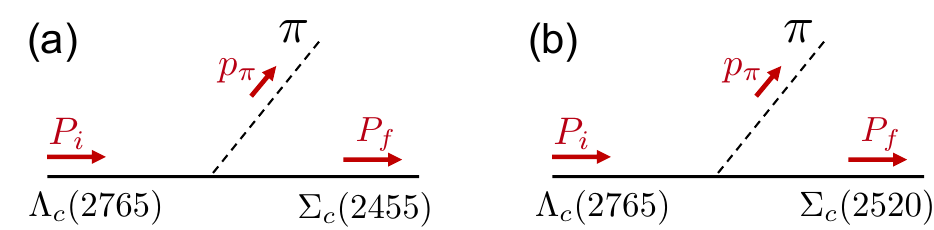}
\caption{Feynman diagrams for (a) $\Lambda_c(2765)\to\Sigma_c(2455)\pi$ and for (b) $\Lambda_c(2765)\to\Sigma_c(2520)\pi$.}
\label{fig:LambdaDecay}
\end{figure}

From Eq.~(\ref{ModelReduced}) together with the spin decomposition in Eq.~(\ref{SpinDec}), transition amplitudes of Fig.~\ref{fig:LambdaDecay} are evaluated as
\begin{eqnarray}
\left|{\cal M}_{1(a)} \right|^2 &=& \frac{2(g_{A,3}^{+H})^2}{3f_0^2}\left|\bar{U}_{f}(P_f) \left(\gamma^\mu-\frac{P_f^\mu}{M_f}\right)\gamma_5p^\pi_\mu U_i(P_i)\right|^2    \nonumber\\
\left|{\cal M}_{1(b)}\right|^2  &=&  \frac{2(g_{A,3}^{+H})^2}{f_0^2}\left|\bar{U}_f^\mu(P_f)p^\pi_\mu U_i(P_i)\right|^2   \ ,  \label{M1ab}
\end{eqnarray}
where ${\cal M}_{1(a)} $ and ${\cal M}_{1(b)}$ correspond to Fig.~\ref{fig:LambdaDecay} (a) and to Fig.~\ref{fig:LambdaDecay} (b), respectively. In Eq.~(\ref{M1ab}), $P^\mu_i$ and $P^\mu_f$ are momenta carried by the heavy baryons of the initial and final states, respectively, and $p^\mu_\pi$ (or $p_\mu^\pi$) is that by the emitted pion as indicated in the figure. The masses of initial- and final-state heavy baryons are denoted by $M_i$ and $M_f$, and the velocity $v^\mu$ has been replaced by $P^\mu_i/M_i$ or $P^\mu_f/M_f$. Besides, $U_i(P_i)$ [$U_f(P_f)$] is the Dirac spinors for the spin-$1/2$ heavy baryon, while $U_f^\mu(P_f)$ is the Rarita-Schwinger spinor for the spin-$3/2$ heavy baryon. That is, those spinors satisfy the following relations~\cite{Falk:1991nq}:
\begin{eqnarray}
&&U_i(P_i)\bar{U}_i(P_i) = \Slash{P}_i+M_i\ , \nonumber\\
&&U_f(P_f)\bar{U}_f(P_f) = \Slash{P}_f+M_f\ , 
\end{eqnarray}
and
\begin{eqnarray}
U_f^\mu(P_f)\bar{U}_f^\nu(P_f) &=& -(\Slash{P}_f+M_f)\Bigg(g^{\mu\nu}-\frac{1}{3}\gamma^\mu\gamma^\nu \nonumber\\
&& -\frac{2}{3M_f^2}P_f^\mu P_f^\nu+\frac{P_f^\mu\gamma^\nu-\gamma^\mu P_f^\nu}{3M_f}\Bigg)\ . \nonumber\\
\end{eqnarray}

Thus, from Eq.~(\ref{M1ab}) the decay widths are evaluated as
\begin{eqnarray}
&&\Gamma[\Lambda_c(2765)\to \Sigma_c(2455)\pi] = \frac{(g_{A,3}^{+H} )^2}{6\pi f_0^2}F_1(M_i,E_f;\pi)\ , \nonumber\\
&&\Gamma[\Lambda_c(2765)\to \Sigma_c(2520)\pi] =\frac{(g_{A,3}^{+H} )^2}{3\pi f_0^2}F_1(M_i,E_f;\pi)\ , \nonumber\\
\label{DecayFormula2}
\end{eqnarray}
for all isospin channels, where we have defined
\begin{eqnarray}
F_1(M_i,E_f;\phi) &\equiv& \frac{(E_f+M_f)|{\bm p}_\phi|}{M_i} \nonumber\\
&\times& \Bigg[ \frac{(E_fE_\phi+|{\bm p}_\phi|^2)^2}{M_f^2}-m_\phi^2\Bigg] \ . 
\end{eqnarray}
In this equation, $E_f=\sqrt{|{\bm P}_f|^2 + M_f^2}$ and $E_\phi = \sqrt{|{\bm p}_\phi|^2+m_\phi^2}$ are dispersion relations of the final-state heavy baryon and of $\phi$, respectively, with $\phi$ representing the emitted NG boson ($m_\phi$ is its mass), in the rest frame of the decaying heavy baryon. The momentum conservation leads to ${\bm p}_\phi = -{\bm P}_f$ and 
\begin{eqnarray}
|{\bm p}_\phi| = \frac{\sqrt{\big[M_i^2-(M_f+m_{\phi})^2\big]\big[M_i^2-(M_f-m_{\phi})^2\big]}}{2M_i} \ . \nonumber\\
\end{eqnarray}
We note that a spin average of the initial-state heavy baryon is taken in obtaining the formulae in Eq.~(\ref{DecayFormula2}). Besides, the coefficients of all isospin channels are identical since $\Lambda_c(2765)$ carries isospin zero which is trivial; the square of Clebsch-Gordan coefficients for all the channels are identical. The formulae show that the analytic form of the decay width of $\Lambda_c(2765)\to\Sigma_c(2520)\pi$ is exactly twice that of $\Lambda_c(2765)\to\Sigma_c(2455)\pi$, which can be understood by the difference of spin degeneracies of the final-state heavy baryon. Such a symmetric form has been provided by the $SU(2)_h$ heavy quark spin symmetry in our present model.

\begin{figure}[tb]
\centering
\includegraphics*[scale=0.53]{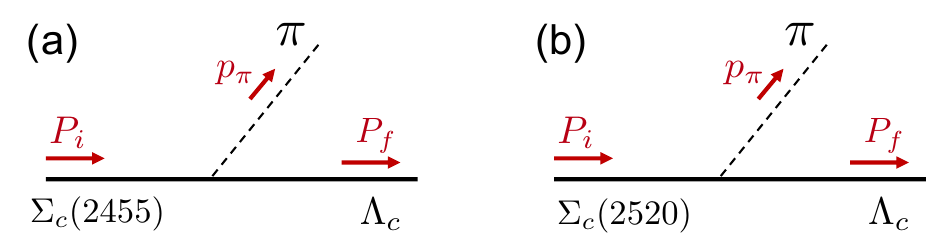}
\caption{Feynman diagrams for (a) $\Sigma_c(2455)\to\Lambda_c\pi$ and for (b) $\Sigma_c(2520)\to\Lambda_c\pi$.}
\label{fig:SigmaDecay}
\end{figure}

Next, from Eq.~(\ref{ModelReduced}) the amplitudes for decays of the HQS-doublet $\{\Sigma_c(2455),\Sigma_c(2520)\}$ are obtained as
\begin{eqnarray}
\left|{\cal M}_{2(a)} \right|^2 &=&  \frac{2(g_{A,3}^{+L})^2}{3f_0^2}\left|\bar{U}_f(P_f)p^{\pi}_\mu \left(\gamma^\mu+\frac{P_i^\mu}{M_i}\right)\gamma_5U_{i}(P_i)\right|^2 \ ,\nonumber\\
\left|{\cal M}_{2(b)}\right|^2  &=&  \frac{2(g_{A,3}^{+L})^2}{f_0^2}\left|\bar{U}_{f}(P_f)p^{\pi}_\mu U_{i}^{\mu}(P_i) \right|^2 \ , \label{M2ab}
\end{eqnarray}
where ${\cal M}_{2(a)}$ and ${\cal M}_{2(b)}$ correspond to Fig.~\ref{fig:SigmaDecay} (a) and Fig.~\ref{fig:SigmaDecay} (b), respectively. The notation should be understood similarly to Eq.~(\ref{M1ab}). Thus, from Eq.~(\ref{M2ab}) the analytic forms of decay widths of $\{\Sigma_c(2455),\Sigma_c(2520)\}$ read
\begin{eqnarray}
&&\Gamma[\Sigma_c(2455)\to\Lambda_c\pi] = \frac{(g_{A,3}^{+L})^2}{6\pi f_0^2}F_2(M_i,E_f;\pi^-)\ , \nonumber\\
&&\Gamma[\Sigma_c(2520)\to\Lambda_c\pi] =  \frac{(g_{A,3}^{+L})^2}{6\pi f_0^2}F_2(M_i,E_f;\pi^-)\ , 
 \label{SigmaDecay2}
\end{eqnarray}
for all isospin channels, with
\begin{eqnarray}
F_2(M_i,E_f;\phi) = \frac{(E_f+M_f)|{\bm p}_\phi|^3}{M_i}\ .
\end{eqnarray}
Equation~(\ref{SigmaDecay2}) shows that the analytic formulae of all the channels are identical because $\Lambda_c$ carries trivial isospin as in Eq.~(\ref{DecayFormula2}). We note that, unlike Eq.~(\ref{DecayFormula2}), the factor for decays of $\Sigma_c(2455)$ and $\Sigma_c(2520)$ coincide due to the $SU(2)_h$ heavy quark spin symmetry, since their spin averages have been taken~\cite{Cho:1994vg}.

The analytic formulae of decay widths of the remaining $\Xi_c$ baryons are obtained in a similar way, by assuming the isospin symmetry. The resultant formulae are summarized in Appendix.~\ref{sec:DecayFormula}


\section{Results}
\label{sec:Results}


 In this section, we investigate decay widths of the Roper-like $\Lambda_c(2765)$, $\Xi_c(2970)$ as well as those of the HQS-doublet $\{\Sigma_c(2455),\Sigma_c(2520)\}$ and $\Xi_c(2645)$ based on the analytic expressions given in Sec.~\ref{sec:Analysis} and in Appendix.~\ref{sec:DecayFormula}. We note that $\Xi_c'$ does not decay by the strong interactions since no thresholds open.

\begin{table}[htbp]
\begin{center}
  \begin{tabular}{cccc} \hline
Baryon & $J^P$ & Mass & Total width  \\ \hline 
$\Lambda_c^+$ & $1/2^+$  & $2286.46$ & No strong decay  \\
$\Xi_c^{+}$ & $1/2^+$  & $2467.95$ & No strong decay   \\
$\Xi_c^{0}$ & $1/2^+$ & $2470.99$ & No strong decay  \\
$\Sigma_c(2455)^{++}$ & $1/2^+$ & $2453.97$  & $1.89$  \\
$\Sigma_c(2455)^{+}$ & $1/2^+$  & $2452.9$ & $<4.6$  \\
$\Sigma_c(2455)^{0}$ & $1/2^+$  & $2453.75$ & $1.83$  \\
$\Sigma_c(2520)^{++}$ & $3/2^+$ & $2518.41$ & $14.78$  \\
$\Sigma_c(2520)^{+}$ & $3/2^+$  & $2517.5$ & $<17$  \\
$\Sigma_c(2520)^{0}$ & $3/2^+$  & $2518.48$ & $15.3$  \\
$\Xi_c'^{+}$ & $1/2^+$  & $2578.2$ & No strong decay   \\
$\Xi_c'^{0}$ & $1/2^+$  & $2578.7$ & No strong decay  \\
$\Xi_c(2645)^{+}$ & $3/2^+$  & $2645.1$ & $2.14$  \\
$\Xi_c(2645)^{0}$ & $3/2^+$ & $2646.16$ & $2.35$  \\
$\Lambda_c(2765)^+$ & $1/2^+$  & $2766.6$ & 50  \\
$\Xi_c(2970)^{+}$ & $1/2^+$  & $2967.1$ & $20.9$  \\
$\Xi_c(2970)^{0}$ & $1/2^+$ & $2965.9$ & $?$ \\
 \hline
 \end{tabular}
\caption{Experimental values of masses and decay widths of the heavy baryons in units of MeV, indicated in the Particle Data Group (PDG)~\cite{Zyla:2020zbs}. In this table only the central values are shown.}
\label{tab:Width}
\end{center}
\end{table}

Before showing the main results, we present properties of the axial charges. Axial charges for negative-parity HQS-singlet baryons $g_{A,i}^{-H}$ and $g_{A,i}^{-L}$ are defined in a similar way to $g_{A,i}^{+H}$ and $g_{A,i}^{+L}$ from the Lagrangian~(\ref{LIntParity}) with the mixing formula~(\ref{Mixing}):
\begin{eqnarray}
g_{A,i}^{-H} &\equiv& g_{A} \sin\theta_{B_-^i}-g_{A}'\cos\theta_{B_-^i} \ , \nonumber\\
g_{A,i}^{-L} &\equiv& g_{A} \cos\theta_{B_-^i} + g_{A}'\sin\theta_{B_-^i} \ .
\end{eqnarray}
Then, from Eq.~(\ref{Couplings}) [and Eq.~(\ref{AxialC1})] we can confirm that the following relation holds:
\begin{eqnarray}
(g_{A,i}^{+H})^2 + (g_{A,i}^{+L})^2  = (g_{A,i}^{-H})^2 + (g_{A,i}^{-L})^2 = R^2\ , \label{GRelation}
\end{eqnarray}
where $R\equiv\sqrt{g_{A}^2+g_{A}'^2}$ represents the magnitude of axial charges, which is independent of the flavor index $i$ due to the $SU(3)_{L+R}$ flavor symmetry. Discussions on Eq.~(\ref{GRelation}) will be given in Sec.~\ref{sec:Discussion}.

Focussing on the positive-parity parts, the relation~(\ref{GRelation}) implies that once $g_{A,i}^{+L}$ is determined, the other $g_{A,i}^{+H}$ can be fixed by the ``radius $R$'' solely regardless of the mixing angles $\theta_{B_+^i}$. Hence, first we fix the parameter $g_{A,i}^{+L}$ by fitting experimentally observed decays of $\{\Sigma_c(2455),\Sigma_c(2520)\}$ and $\Xi_c(2645)$ listed in Table~\ref{tab:Width}. And next, with the help of Eq.~(\ref{GRelation}) we examine decay widths of the Roper-like $\Lambda_c(2765)$ and $\Xi_c(2970)$ by varying $R$.

Owing to the $SU(2)_h$ heavy quark spin symmetry, the axial charges contributing to decays of the HQS-doublet baryons are common among the spin partners. Based on this fact, using the experimental values in Table~\ref{tab:Width}, we evaluate the axial charges $g_{A,i}^{+L}$ by the following averages:
\begin{eqnarray}
g_{A,3}^{+L} &=&\frac{1}{4} \Big( g_{A,3}^{+L}\Big|_{\Sigma_c(2455)^{++}} + g_{A,3}^{+L}\Big|_{\Sigma_c(2455)^{0}} \nonumber\\
&& + g_{A,3}^{+L}\Big|_{\Sigma_c(2520)^{++}} + g_{A,3}^{+L}\Big|_{\Sigma_c(2520)^{0}}\Big) \nonumber\\
&=&0.512\ , \label{GTilde1}
\end{eqnarray}
and
\begin{eqnarray}
g_{A,1}^{+L} &=& \frac{1}{2}\Bigg(g_{A,1}^{+L}\Big|_{\Xi_c(2645)^+} + g_{A,1}^{+L}\Big|_{\Xi_c(2645)^0}\Bigg)  \nonumber\\
&=& 0.511\ , \label{GTilde2}
\end{eqnarray}
where $g_{A,i}^{+L}>0$ has been assumed.\footnote{In the non-relativistic quark model, those axial charges are expected to be about $1$. Hence, in order to reproduce the axial charges of about $0.5$ as in Eq.~(\ref{GTilde1}) [Eq.~(\ref{GTilde2})] in the quark model, we need to include a phenomenological suppression parameter by hand~\cite{Yan:1992gz}. We also note that such a suppression is systematically realized by including relativistic corrections to the non-relativistic quark model~\cite{Arifi:2021orx}. } In Eq.~(\ref{GTilde1}), we have not included the widths of $\Sigma_c(2455)^+$ and $\Sigma_c(2520)^+$ to estimate $g_{A,3}^{+L}$ since only the upper limit is known as indicated in Table~\ref{tab:Width}. Also, in Eq.~(\ref{GTilde2}) only decays of $\Xi_c(2645)$ are taken into account since $\Xi_c'$ does not decay, and the isospin symmetry has been taken: $g_{A,1}^{+L} =g_{A,2}^{+L}$. Interestingly, Eqs~(\ref{GTilde1}) and~(\ref{GTilde2}) show that the $SU(3)_{L+R}$ flavor symmetry is realized as a good approximation for the axial charges $g_{A,3}^{+L}$ and $g_{A,1}^{+L}$. Therefore, in what follows, we assume the $SU(3)_{L+R}$ flavor symmetry for the axial charges to satisfy $g_{A,3}^{+L}=g_{A,1}^{+L} \equiv g_{A}^{+L}$, where 
\begin{eqnarray}
g_{A}^{+L}= \frac{1}{2}\Big( g_{A,3}^{+L}+ g_{A,1}^{+L} \Big) = 0.512\ .
\end{eqnarray}
It should be noted that such an assumption leads to $\theta_{B_{+}^1} = \theta_{B_{+}^2} = \theta_{B_{+}^3} \equiv \theta_{B_+}$ for the mixing angle defined in Eq.~(\ref{Mixing}).

\begin{figure}[tb]
\centering
\includegraphics*[scale=0.55]{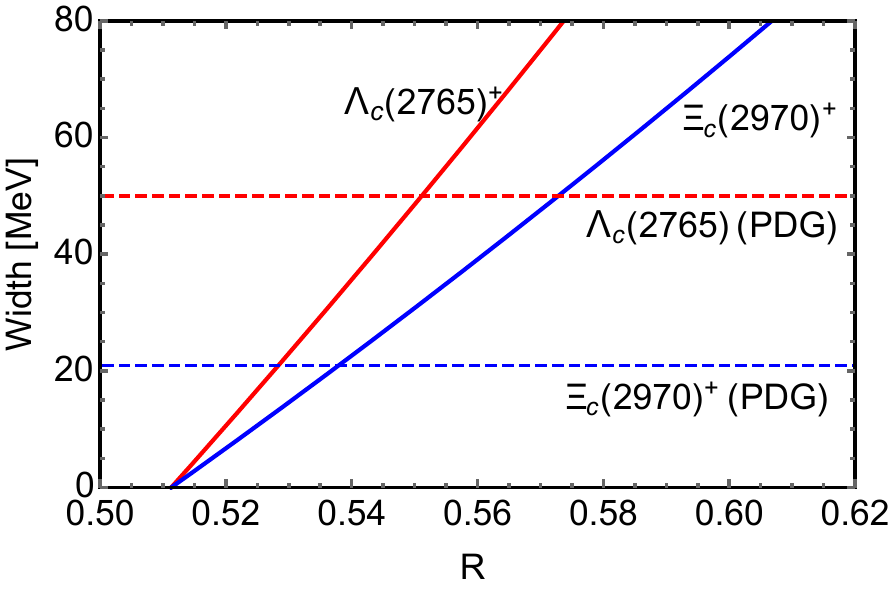}
\caption{$R=\sqrt{g_{A}^2+g_{A}'^2}$ dependence of the decay widths of $\Lambda_c(2765)$ (red) and $\Xi_c(2970)^+$ (blue). The dashed lines are the corresponding experimental values (the central values) indicated in PDG.}
\label{fig:RoperWidth}
\end{figure}

\begin{table}[htbp]
\begin{center}
  \begin{tabular}{cccc} \hline
Decaying baryon & Channel & Width & (PDG value) \\ \hline 
$\Lambda_c(2765)^+$ & $\Sigma_c(2455)\pi$ & $12.7$ - $41.6$ \\
 & $\Sigma_c(2520)\pi$ & $10.2$ - $33.4$ \\
  & Sum & $23.0$ - $75.0$ & ($50$)  \\ \hline
$\Xi_c(2970)^+$ & $\Sigma_c(2455)\bar{K}$ & $0.704$ - $2.30$ \\
 & $\Xi_c'\pi$ & $6.83$ - $22.3$ \\
 & $\Xi_c(2645)$ & $7.03$ - $22.9$ \\
  & Sum & $14.6$ - $47.6$ & ($20.9^{+2.4}_{-3.5}$) \\ \hline
 $\Xi_c(2970)^0$ & $\Sigma_c(2455)\bar{K}$ & $0.577$ - $1.88$ \\
 & $\Xi_c'\pi$ & $6.78$ - $22.1$ \\
 & $\Xi_c(2645)$ & $6.97$ - $22.7$ \\
 & Sum & $14.3$ - $46.7$ \\ \hline
 \end{tabular}
\caption{Partial and total decay widths of $\Lambda_c(2765)$ and $\Xi_c(2970)$ for $R = 0.53$ - $0.57$ in units of MeV.}
\label{tab:WidthSummary}
\end{center}
\end{table}

In Fig.~\ref{fig:RoperWidth}, the resultant $R=\sqrt{g_A^2+g_A'^2}$ dependence of decay widths of $\Lambda_c(2765)^+$ (red) and $\Xi_c(2970)^+$ (blue) is plotted. In this figure, the red and blue dashed lines are the experimental values of $\Lambda_c(2765)^+$ and $\Xi_c(2970)^+$ indicated in PDG~\cite{Zyla:2020zbs}, respectively. Figure~\ref{fig:RoperWidth} shows that the decay widths of $\Lambda_c(2765)^+$ and $\Xi_c(2970)^+$ whose magnitudes are tens of MeV, which cannot be explained by the conventional non-relativistic quark model~\cite{Nagahiro:2016nsx}, are reasonably obtained for $ R \sim 0.53$ - $0.57$. In the figure, we have not shown the result for $\Xi_c(2970)^0$ since PDG does not indicate the decay width explicitly. 
For convenience, in Table~\ref{tab:WidthSummary}, we show partial and total decay widths of $\Lambda_c(2765)$ and $\Xi_c(2970)$ for $ R = 0.53$ - $0.57$. From the table, one can see that the decay widths are very sensitive to the value of $R$, i.e., the magnitude of axial charges.

When we choose the value of $R$ to be $R=0.55$, the resultant axial charge $g_{A}^{+H}$ reads $g_A^{+H}=0.201$ (with an assumption of $g_{A}^{+H}>0$). To summarize, the magnitude of axial charges related to $\Lambda_c(2765)$-$\{\Sigma_c(2455),\Sigma_c(2520)\}$ sector and $\Lambda_c$-$\{\Sigma_c(2455),\Sigma_c(2520)\}$ sector are $g_{A}^{+H}\approx 0.201$ and $g_{A}^{+L}=0.512$, respectively. 

The above analysis has been done with $f_0=\frac{1}{2}(f_\pi+f_K)=101\, {\rm MeV}$. When we use $f_0=f_\pi=92.1\, {\rm MeV}$ the axial charge reads $g_{A}^{+L}=0.466$, and the value of $R$ at which $\Gamma_{\Lambda_c(2765)}=50\, {\rm MeV}$ becomes $R=0.502$ [this value is $R=0.551$ for $f_0=\frac{1}{2}(f_\pi+f_K)=101\, {\rm MeV}$]. Besides, when we use $f_0=\frac{1}{3}(f_\pi+f_K+f_\eta)=107\, {\rm MeV}$ the axial charge reads $g_{A}^{+L}=0.543$, and the value of $R$ at which $\Gamma_{\Lambda_c(2765)}=50\, {\rm MeV}$ becomes $R=0.585$. Those demonstrations show that the axial charges $g_A^{L+}$ and $R$ do not change significantly for other estimates of $f_0$.

\begin{table}[htbp]
\begin{center}
  \begin{tabular}{cccc} \hline
Baryon & $J^P$ & Mass & Total width \\ \hline 
$\Lambda_b^0$ & $1/2^+$  & $5619.60$ & No strong decay  \\
$\Xi_b^{-}$ & $1/2^+$  & $5797.0$ & No strong decay   \\
$\Xi_b^{0}$ & $1/2^+$ & $5791.9$ & No strong decay  \\
$\Sigma_b^{+}$ & $1/2^+$ & $5810.56$  & $5.0$  \\
$\Sigma_b^{0}$ &$1/2^+$ & ? & ? \\
$\Sigma_b^{-}$ & $1/2^+$  & $5815.64$ & $5.3$  \\
$\Sigma_b^{*+}$ & $3/2^+$ & $5830.32$ & $9.4$  \\
$\Sigma_b^{*0}$ &$3/2^+$ & ? & ? \\
$\Sigma_b^{*-}$ & $3/2^+$  & $5834.74$ & $10.4$  \\
$\Xi_b'^{0}$ & $1/2^+$  & ? & ?   \\
$\Xi_b'(5935)^{-}$ & $1/2^+$  & $5935.02$ & $<0.08$   \\
$\Xi_b(5945)^{0}$ & $3/2^+$  & $5952.3$ & $0.90$  \\
$\Xi_b(5955)^{-}$ & $3/2^+$ & $5955.33$ & $1.65$  \\
$\Lambda_b(6070)^0$ & $1/2^+$  & $6072.3$ & 72  \\
 \hline
 \end{tabular}
\caption{Experimental values of masses and decay widths of the bottom baryons in units of MeV, indicated in PDG~\cite{Zyla:2020zbs}. In this table only the central values are shown. We note that the spin and parity of $\Lambda_b(6070)$ are assumed to be $J^P=1/2^+$ as suggests by the recent experimental data~\cite{CMS:2020zzv,Aaij:2020rkw}.}
\label{tab:WidthBottom}
\end{center}
\end{table}

\section{Bottom baryons}
\label{sec:Bottom}

In this section, we investigate bottom baryons in a similar way to Sec.~\ref{sec:Results}. The experimentally observed masses and widths of the relevant bottom baryons are listed in  Table~\ref{tab:WidthBottom}.

As done in Sec.~\ref{sec:Results}, we determine $g_{A,i}^{+L}$ for bottom baryons by fitting decay widths of the HQS-doublet $\Sigma_b$'s and $\Xi_b$'s. Using the available experimental values in Table~\ref{tab:WidthBottom}, we can get the following averaged values:
\begin{eqnarray}
g_{A,3}^{+L} &=&\frac{1}{4} \Big(g_{A,3}^{+L}\Big|_{\Sigma_b^+} + g_{A,3}^{+L}\Big|_{\Sigma_b^-}  +  g_{A,3}^{+L}\Big|_{\Sigma_b^{*+}} + g_{A,3}^{+L}\Big|_{\Sigma_b^{*-}}\Big) \nonumber\\
&=& 0.488 \ , \label{GTildeB1}
\end{eqnarray}
and
\begin{eqnarray}
g_{A,1}^{+L}&=& \frac{1}{2}\Big(g_{A,1}^{+L}\Big|_{\Xi_b(5945)^0} + g_{A,1}^{+L}\Big|_{\Xi_b(5955)^-}\Big)  \nonumber\\
&=& 0.577 \ , \label{GTildeB2}
\end{eqnarray}
together with the isospin symmetry, for the bottom baryons. Similarly to the charmed baryons, we have assumed $g_{A,i}^{+L}>0$. Equations~(\ref{GTildeB1}) and~(\ref{GTildeB2}) imply that the violation of $SU(3)_{L+R}$ flavor symmetry for bottom baryons is slightly larger than that for charmed baryons, but still the deviation is small. Hence, again we assume the $SU(3)_{L+R}$ flavor symmetry as $g_{A,3}^{+L}=g_{A,1}^{+L}\equiv g_{A}^{+L}$, with $g_{A}^{+L}$ being estimated as
\begin{eqnarray}
g_{A}^{+L} =  \frac{1}{2}\Big(g_{A,3}^{+L} + g_{A,1}^{+L} \Big) = 0.532 \ . \label{GA+LBottom}
\end{eqnarray}
In this estimate the deviation reads
\begin{eqnarray}
\delta g_A^{+L} \equiv \frac{g_A^{+L}-g_{A,3}^{+L}}{g_A^{+L}} = \frac{g_{A,1}^{+L}-g_{A}^{+L}}{g_A^{+L}} = 8.37\, \%\ ,
\end{eqnarray}
which is still less than ten percent.

In determining $g_{A}^{+L}$, we have employed the central values of decay widths indicated in PDG whose error bars are larger  for the bottom sector than for the charm sector. Thus, it is expected that the $SU(3)_{L+R}$ flavor symmetry works better for the bottom baryons like the charmed baryons when we choose other values within the error bars. However, such precise determination of $g_{A,i}^{+L}$ is not our main aim, and we will employ Eq.~(\ref{GA+LBottom}) in the following analysis.

A candidate for the Roper-like $\Lambda_b$ is $\Lambda_b(6070)$ as indicated in Table~\ref{tab:WidthBottom}, while the Roper-like $\Xi_b$ has not been experimentally observed. Thus, here we fix the value of $R$ for bottom baryons by fitting the width of $\Lambda_b(6070)$ from $\Lambda_b(6070)\to\Sigma_b\pi$ and $\Lambda_b(6070)\to\Sigma_b^*\pi$ channels. The resultant value is 
\begin{eqnarray}
R= 0.612 \ . \label{HBottom}
\end{eqnarray}
We note that the unknown masses of $\Sigma_b^0$ and $\Sigma_b^{*0}$ have been estimated by averaging masses of available isospin partners: $M_{\Sigma_b^0} = \frac{1}{2}(M_{\Sigma_b^+}+M_{\Sigma_b^-}) = 5813.1$ MeV, and $M_{\Sigma_b^{*0}} = \frac{1}{2}(M_{\Sigma_b^{*+}}+M_{\Sigma_b^{*-}}) = 5832.53$ MeV, in obtaining Eq.~(\ref{HBottom}). $R$ for the charmed baryons has been considered to be $R\sim 0.53$ - $0.57$, thus Eq.~(\ref{HBottom}) shows that the heavy quark flavor symmetry for the axial charges seems to be satisfied well. The value of $g_A^{+H}$ for bottom baryons are estimated as $g_A^{+H}=0.303$ with $g_A^{+H}>0$ assumed.

One theoretical candidate for the undiscovered Roper-like $\Xi_b$ is the $\Xi_b(6255)$ predicted in Ref.~\cite{Chen:2018orb}. Therefore, we employ the mass of $\Xi_b(6255)$ as that of the Roper-like $\Xi_b$, and examine its partial and total widths together with $\Lambda_b(6070)$ ones. The results are summarized in Table~\ref{tab:WidthBSummary}. In this table we have used $R= 0.612 $, and assumed that the unknown mass of $\Xi_b'^0$ is identical to that of $\Xi_b'(5935)^-$ from isospin symmetry.

\begin{table}[htbp]
\begin{center}
  \begin{tabular}{cccc} \hline
Decaying baryon & Channel & Width & (PDG value)  \\ \hline 
$\Lambda_b(6070)^0$ & $\Sigma_b\pi$ & $29.6$ \\
 & $\Sigma_b^*\pi$ & $42.4$ \\
  & Sum & $72$ & ($72\pm11\pm2$) \\ \hline
$\Xi_b(6255)^0$ &  $\Xi_b'\pi$ & $16.9$ \\
 & $\Xi_b'^*\pi$ & $26.7$ \\
  & Sum & $43.5$ \\ \hline
$\Xi_b(6255)^-$ &  $\Xi_b'\pi$ & $16.8$ \\
 & $\Xi_b'^*\pi$ & $27.0$ \\
  & Sum & $43.8$ \\ \hline
 \end{tabular}
\caption{Partial and total decay widths of $\Lambda_b(6070)$ and $\Xi_b(6255)$ for $R= 0.612 $ in units of MeV. In this table $\Xi_b'$ and $\Xi_b'^*$ collectively denote $\{\Xi_b'^0,\Xi_b'(5935)^-\}$ and $\{\Xi_b(5945)^0,\Xi_b(5955)^-\}$, respectively.}
\label{tab:WidthBSummary}
\end{center}
\end{table}

\begin{figure}[tb]
\centering
\includegraphics*[scale=0.55]{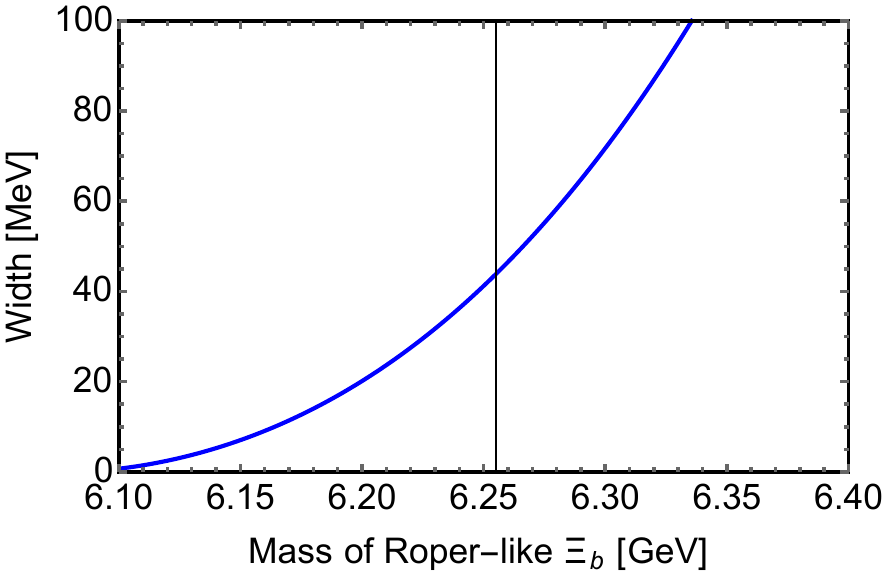}
\caption{ Mass dependence of the decay width of Roper-like $\Xi_b$. The vertical black line represents the mass of $\Xi_b(6255)$ predicted in Ref.~\cite{Chen:2018orb}. }
\label{fig:XiRoperMassDep}
\end{figure}

\begin{figure}[tb]
\centering
\includegraphics*[scale=0.57]{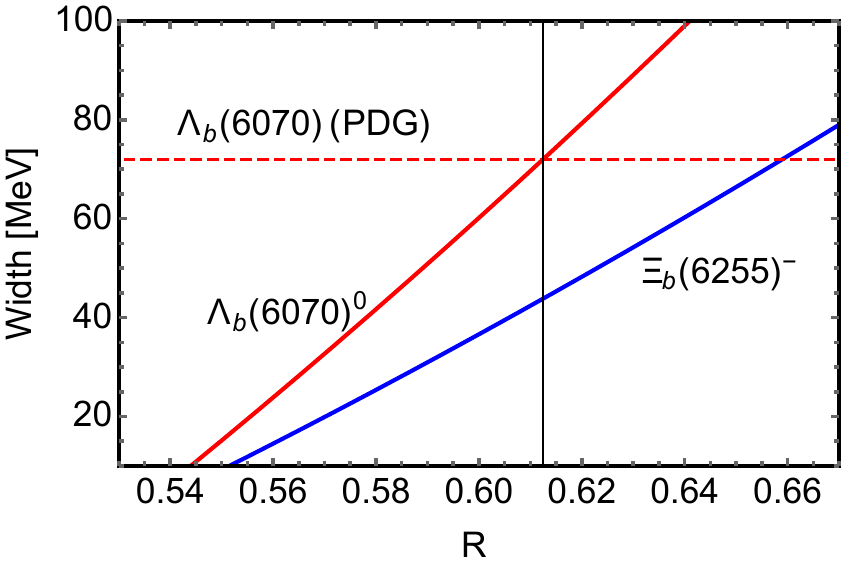}
\caption{$R=\sqrt{g_{A}^2+g_{A}'^2}$ dependence of the decay widths of $\Lambda_b^0(6070)$ (red) and the Roper-like $\Xi_b^-$ (blue). The red dashed line is the experimental values (the central value) indicated in PDG, and the vertical black line corresponds to $R=0.612$. }
\label{fig:RoperWidthB}
\end{figure}

In order to examine other possibilities of the Roper-like $\Xi_b$, we show the mass dependence of its decay width in Fig.~\ref{fig:XiRoperMassDep} for $R=0.612$. In this figure the vertical black line represents the mass of $\Xi_b(6255)$ predicted in Ref.~\cite{Chen:2018orb}, corresponding to the values indicated in Table~\ref{tab:WidthBSummary}. We have shown only the width of charged $\Xi_b^-$, since the results for $\Xi_b^0$ is almost identical to the $\Xi_b^-$ one. Figure~\ref{fig:XiRoperMassDep} indicates that the decay width of Roper-like $\Xi_b$ is expected to be tens of MeV, when we assume its mass is larger than the mass of ground-state $\Xi_b$ by about $500$ MeV ($\sim 6.3$ GeV) as naively anticipated. Depicted in Fig.~\ref{fig:RoperWidthB} is $R$ dependence of the decay width of Roper-like $\Lambda_b(6070)$ (red) and theoretically predicted $\Xi_b(6255)$ (blue). The red dashed horizontal line indicates the PDG value for the decay width of $\Lambda_b(6070)$, and the vertical black line corresponds to $R=0.612$ in Eq.~(\ref{HBottom}). Figures~\ref{fig:XiRoperMassDep} and~\ref{fig:RoperWidthB} provide future experiments with useful information on the decay width of the undiscovered Roper-like $\Xi_b$.

\section{Discussions}
\label{sec:Discussion}

\subsection{Discussions on the relation~(\ref{GRelation})}
\label{sec:AxialRelation}

In this subsection we discuss the relation satisfied by axial charges of the heavy baryons presented in Eq.~(\ref{GRelation}). This relation is one of the unique features of our present model, controlling couplings among the heavy baryons and the light mesons. Here, we summarize its properties:
\begin{itemize}
\item The relation holds regardless of the mixing angles for the HQS-singlet baryons in Eq.~(\ref{Mixing}). In other words, the relation is derived only by a fact that the ground-state and the Roper-like heavy baryons are described by mixings of the components~(\ref{BRBLFields}), and information on the detailed mixing properties is not essential.
\item The relation does not include the flavor index $i$, so that it links axial charges for the different flavor sectors with the identical magnitude, as long as $SU(3)_{L+R}$ flavor symmetry works well as explicitly shown in the present investigation.
\item Moreover, the relation includes the axial charges associated with the negative-parity HQS-singlet baryons in a similar way to those with the positive-parity ones, sharing the identical magnitude $R$. This is a significant consequence of employing the chiral model, and is expected to play an important role in pursuing the parity partners~\cite{Harada:2019udr,Kawakami:2020sxd} as will be further discussed in Sec.~\ref{sec:ParityPartner}.
\end{itemize}

In the remainder of this subsection, we discuss higher-order corrections to the relation~(\ref{GRelation}). In Sec.~\ref{sec:Lagrangian} we have constructed the interaction Lagrangian for the heavy baryons and a NG boson with one $\partial_\mu\Sigma$ or $\partial_\mu\Sigma^\dagger$. As for one NG boson emission decays of the heavy baryons, momentum carried by the NG boson is of ${\cal O}(\Lambda_{\rm QCD})$ which is small  compared to heavy baryon masses. For this reason, our truncation that only one $\partial_\mu\Sigma$ or $\partial_\mu\Sigma^\dagger$ is left seems to work well based on the spirit of heavy quark symmetry. 

Apart from the above corrections, it may be possible to include higher-oder corrections with respect to non-derivative terms of $\Sigma$, $\Sigma^\dagger$, $\Sigma^T$ and $\Sigma^{*}$ to Eq.~(\ref{LInt}). For instance, 
 \begin{eqnarray}
{\cal L}^{\rm higher}_{\rm int} &=&  \frac{4\sqrt{2}\tilde{g}_A}{f_0^3} \nonumber\\
&\times&  {\rm Tr}\Big[\bar{\cal B}_R(\partial_\mu\Sigma^\dagger)\Sigma \Sigma^\dagger S^{T\mu} + \bar{\cal B}_L(\partial_\mu\Sigma)\Sigma^\dagger\Sigma S^\mu +  {\rm h.c.}\Big] \nonumber\\
&+&  \frac{4\sqrt{2} \tilde{g}_A' }{f_0^3}\nonumber\\ 
&\times& {\rm Tr}\Big[\bar{\cal B}_L'(\partial_\mu\Sigma^\dagger)\Sigma\Sigma^\dagger S^{T\mu}+\bar{\cal B}_R'(\partial_\mu\Sigma)\Sigma^\dagger\Sigma S^\mu +  {\rm h.c.}\Big]  \nonumber\\
&+& \cdots \ ,  
\label{ImpInt}
 \end{eqnarray}
is possible as one of the simplest next-to-leading oder contributions. We note that the even number of $\Sigma$, $\Sigma^\dagger$, $\Sigma^T$ and $\Sigma^{*}$ with one derivatives are forbidden due to chiral symmetry. Under the spontaneous breakdown of chiral symmetry the $\Sigma$ is replaced by its VEV: $\langle\Sigma\rangle = \frac{f_0}{2}{\bm 1}$ when its fluctuations are neglected, and thus the Lagrangian~(\ref{ImpInt}) turns into
\begin{eqnarray}
{\cal L}^{\rm higher}_{\rm int} &=& \frac{\sqrt{2}\tilde{g}_A }{f_0}  {\rm Tr}\Big[\bar{\cal B}_R\partial_\mu\Sigma^\dagger S^{T\mu} + \bar{\cal B}_L\partial_\mu\Sigma S^\mu +  {\rm h.c.}\Big] \nonumber\\
&+&  \frac{\sqrt{2}\tilde{g}_A' }{f_0}  {\rm Tr}\Big[\bar{\cal B}_L'\partial_\mu\Sigma^\dagger S^{T\mu}+\bar{\cal B}_R'\partial_\mu\Sigma  S^\mu +  {\rm h.c.}\Big]  \nonumber\\
&+& \cdots \ ,  
\label{ImpInt2}
 \end{eqnarray}
by abbreviating multi NG boson interactions. Apparently Eq.~(\ref{ImpInt2}) exhibits the same structure as in Eq.~(\ref{LInt}). The structure has arisen since the VEV of $\Sigma$ is of the form $\frac{f_0}{2}{\bm 1}$ proportional to the unit matrix, and such a coincidence is also true for any higher-order contributions. Therefore, argument for $g_A$ and $g_A'$ below Eq.~(\ref{LInt}) can be applied here in a similar way; the couplings incorporating higher-order contributions defined by
\begin{eqnarray}
G_A \equiv g_A + \tilde{g}_A + \cdots\ , \ \ G'_A \equiv g'_A + \tilde{g}'_A + \cdots\ ,
\end{eqnarray}
coincide with the axial charges as derived in Eq.~(\ref{LAxial}), and the relation for the axial charges shown in Eq.~(\ref{GRelation}) is found to hold at any higher order as
\begin{eqnarray}
(G_{A,i}^{+H})^2 + (G_{A,i}^{+L})^2 = (G_{A,i}^{-H})^2 + (G_{A,i}^{-L})^2 = {\cal R}^2 \ , \label{GRelationH}
\end{eqnarray}
with ${\cal R}=\sqrt{G_A^2+G_A'^2}$ and
\begin{eqnarray}
G_{A,i}^{\pm H} &\equiv& G_{A} \sin\theta_{B_\pm^i}-G_{A}'\cos\theta_{B_\pm^i} \ , \nonumber\\
G_{A,i}^{\pm L} &\equiv& G_{A} \cos\theta_{B_\pm^i} + G_{A}'\sin\theta_{B_\pm^i} \ .\label{CouplingsH}
\end{eqnarray}

The above demonstration implies that the relation~(\ref{GRelationH}) is derived simply as consequences of the inclusion of two HQS-singlet baryons, belonging to (anti-)fundamental representations of $SU(3)_L\times SU(3)_R$ as in Eqs.~(\ref{ChiralDiquark1}) and~(\ref{ChiralDiquark2}), which are necessary to describe both the Roper-like as well as the ground-state singly heavy baryons. 



\subsection{Decays of the parity partners}
\label{sec:ParityPartner}

In addition to the positive-parity heavy baryons examined in this article, pursuing the undiscovered negative-parity ones, whose existences are naturally predicted as parity partners~\cite{Harada:2019udr,Kawakami:2020sxd}, are of great importance toward elucidation of chiral symmetric properties of the heavy baryons. In this subsection we comment on predictions in association with such partners.

The axial charges associated with the negative-parity HQS-singlet baryons satisfy the relation~(\ref{GRelation}) [or more generally Eq.~(\ref{GRelationH})] sharing the identical magnitude $R$ to those with the positive-parity ones. Those structures are one of the notable features provided by the parity partners in a chiral model with the linear representation. Hence, our present model can predict decay widths where $S_{\bar{3}}^\mu$ is involved without introducing additional parameters.

The $S_{\bar{3}}^\mu$ denotes the HQS-doublet baryons with $J^P=1/2^-,3/2^-$ belonging to the flavor anti-symmetric representation. For charmed baryons, $S_{\bar{3}}^\mu$ contains $\{\Lambda_c(2595),\Lambda_c(2625)\}$ and $\{\Xi_c(2790),\Xi_c(2815)\}$ indicated in PDG. Thus, the interaction Lagrangian~(\ref{LIntParity}) can predict the decay widths of one NG boson emission processes such as $\Xi_c(1/2^-)\to \Xi_c(2790)\pi$, etc., in principle when the threshold opened.\footnote{Here, $\Xi_c(1/2^-)$ is the undiscovered parity partner to the ground-state $\Xi_c$. The negative-parity $\Lambda_c(1/2^-)$ is also defined in the same way.} Detailed estimation of such partial decay widths requires understandings of the masses of $\Xi_c(1/2^-)$, in addition to our finding that the magnitude of axial charges is $R\approx0.55$ in Sec.~\ref{sec:Results}.

Moreover, for the decay of $\Lambda_c(1/2^-)$, as argued in Ref.~\cite{Kawakami:2020sxd} the $S$-wave decay of $\Lambda_c(1/2^-)\to\Lambda_c\eta$ would not be neglected if the threshold is opened, and this may be true for decays of $\Xi_c(1/2^-)$. These decays cannot be described by our present model~(\ref{LIntParity}), implying that extension of our present model is inevitable. The main decay modes are largely dependent on the masses of $\Lambda_c(1/2^-)$ and $\Xi_c(1/2^-)$, and we leave those examinations on the undiscovered parity partners to future study.

\section{Summary}
\label{sec:Conclusions}

In this paper, we have investigated decay properties of the Roper-like singly heavy baryons $\Lambda_c(2765)$ and $\Xi_c(2970)$, as well as those of the HQS-doublet ones $\{\Sigma_c(2455),\Sigma_c(2520)\}$ and $\{\Xi_c',\Xi_c(2645)\}$ within a chiral model. Based on the model, we have derived a relation satisfied by axial charges of the heavy baryons that controls magnitude of their one pion (kaon) decays. Employing reasonable values of the axial charges, our chiral model has successfully explained large decay widths of the Roper-like $\Lambda_c(2765)$ and $\Xi_c(2970)$. In addition to the investigation of charmed baryons, we have applied our model to bottom baryons to estimate a decay width of the undiscovered Roper-like $\Xi_b$, predicting tens of MeV.

We expect that our present investigation leads to elucidation of dynamical properties of the excited heavy baryons from the viewpoint of chiral symmetry. In particular, the relation that axial charges of the heavy baryons satisfy [Eq.~(\ref{GRelation})] is one of the unique features of making use of a chiral model together with the $SU(2)_h$ heavy quark spin symmetry, which cannot be derived by quark models.
Besides, the estimated value of the axial charges is expected to be a benchmark for the investigation of the heavy baryons in the future. In addition, we emphasize that pursuing the undiscovered parity partners of the ground-state $\Lambda_c$ and $\Xi_c$ is indispensable for further unveiling chiral structures of the singly heavy baryons.

\section*{acknowledgement}

D.S. thanks D. Jido for useful comments on the mirror diquarks and A. J. Arifi for comments on the suppression of direct decays of the Roper-like $\Lambda_c(2765)$. A.H. was supported in part by Grants-in Aid for Scientific Research on Innovative Areas (No. 18H05407).

\begin{figure*}[tb]
\centering
\includegraphics*[scale=0.65]{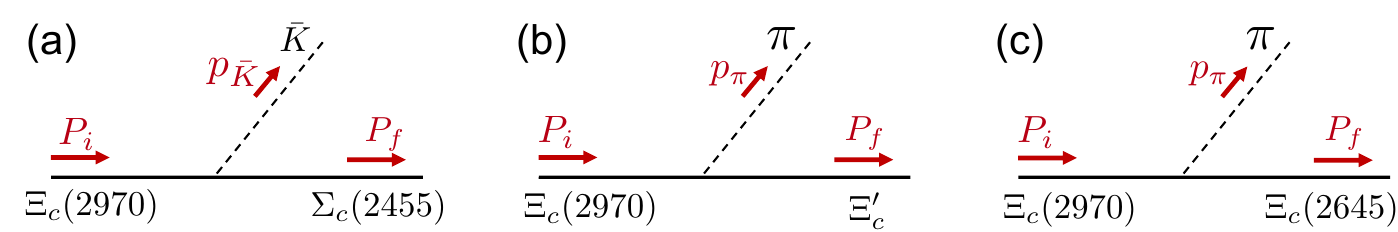}
\caption{Feynman diagrams for (a) $\Xi_c(2970)\to\Sigma_c(2455)\bar{K}$, (b) $\Xi_c(2970)\to\Xi_c'\pi$, and (c) $\Xi_c(2970)\to\Xi_c(2645)\pi$.}
\label{fig:XiDecay}
\end{figure*}

\begin{figure}[tb]
\centering
\includegraphics*[scale=0.7]{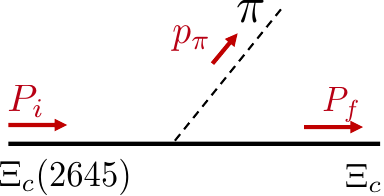}
\caption{A Feynman diagram for $\Xi_c(2645)\to\Xi_c\pi$.}
\label{fig:XiC1Decay}
\end{figure}

\appendix

\section{Analytic expressions of decay widths of $\Xi_c(2970)$ and $\Xi_c(2645)$}
\label{sec:DecayFormula}

In this appendix we summarize analytic expressions of decay widths of the Roper-like $\Xi_c(2970)$, and those of the HQS-doublet $\Xi_c(2645)$. The relevant diagrams are illustrated in Figs.~\ref{fig:XiDecay}. We note that $\Xi_c'$ does not decay by the strong interactions since no thresholds open.

In the following calculations, we will assume isospin symmetry for the mixing angles: $\theta_{B_{+}^1} = \theta_{B_{+}^2}$, and hence we will focus on only $i=1$. As in Eqs.~(\ref{ModelReduced}) and~(\ref{Couplings}), the decay widths are controlled by axial charges defined by
\begin{eqnarray}
g_{A,1}^{+H} &\equiv& g_A \sin\theta_{B_+^1}-g_A'\cos\theta_{B_+^1} \ , \nonumber\\
g_{A,1}^{+L} &\equiv& g_A \cos\theta_{B_+^1} + g_A'\sin\theta_{B_+^1} \ . \label{AxialC1}
\end{eqnarray}
Here, it should be noted that $\Xi_c$ baryons carry isospin-$1/2$, so that coefficients for decay widths in different isospin channels can vary. For this reason we present all the decay channels explicitly. Following the procedure explained in Sec.~\ref{sec:Analysis}, the decay widths for the Roper-like $\Xi_c(2970)$ are evaluated as
\begin{eqnarray}
&&\Gamma[\Xi_c(2970)^0\to \Sigma_c(2455)^+{K}^-] = \frac{(g_{A,1}^{+H})^2}{12\pi f_0^2} \nonumber\\
&& \hspace{5.2cm} \times F_1(M_i,E_f;{K}^-) \ , \nonumber\\
&& \Gamma[\Xi_c(2970)^0\to \Sigma_c(2455)^0\bar{K}^0] =  \frac{(g_{A,1}^{+H})^2}{6\pi f_0^2}F_1(M_i,E_f;\bar{K}^0)\ ,  \nonumber\\
&&\Gamma[\Xi_c(2970)^+\to \Sigma_c(2455)^{++}{K}^-] = \frac{(g_{A,1}^{+H})^2}{6\pi f_0^2} \nonumber\\
&& \hspace{5.2cm} \times F_1(M_i,E_f;{K}^-) \ , \nonumber\\
&& \Gamma[\Xi_c(2970)^+\to \Sigma_c(2455)^{+}\bar{K}^0] =  \frac{(g_{A,1}^{+H})^2}{12\pi f_0^2}F_1(M_i,E_f;\bar{K}^0) \ , \nonumber\\
\end{eqnarray}
for Fig.~\ref{fig:XiDecay} (a), 
\begin{eqnarray}
&&\Gamma[\Xi_c(2970)^0\to \Xi_c'^+\pi^-] = \frac{(g_{A,1}^{+H})^2}{12\pi f_0^2} F_1(M_i,E_f;\pi^-)\ , \nonumber\\
&&\Gamma[\Xi_c(2970)^0\to \Xi_c'^{0}\pi^0] = \frac{(g_{A,1}^{+H})^2}{24\pi f_0^2}F_1(M_i,E_f;\pi^0) \ ,\nonumber\\
&&\Gamma[\Xi_c(2970)^+\to \Xi_c'^0\pi^+] =  \frac{(g_{A,1}^{+H})^2}{12\pi f_0^2}F_1(M_i,E_f;\pi^+) \ ,  \nonumber\\
&& \Gamma[\Xi_c(2970)^+\to \Xi_c'^{+}\pi^0] =  \frac{(g_{A,1}^{+H})^2}{24\pi f_0^2}F_1(M_i,E_f;\pi^0) \, ,
\end{eqnarray}
for Fig.~\ref{fig:XiDecay} (b), and
\begin{eqnarray}
&&\Gamma[\Xi_c(2970)^0\to \Xi_c(2645)^+\pi^-] =  \frac{(g_{A,1}^{+H})^2}{6\pi f_0^2}F_1(M_i,E_f;\pi^-)\ , \nonumber\\
&&\Gamma[\Xi_c(2970)^0\to \Xi_c(2645)^0\pi^0] = \frac{(g_{A,1}^{+H})^2}{12\pi f_0^2}F_1(M_i,E_f;\pi^0)\ , \nonumber\\
&& \Gamma[\Xi_c(2970)^+\to \Xi_c(2645)^0\pi^+] =  \frac{(g_{A,1}^{+H})^2}{6\pi f_0^2}F_1(M_i,E_f;\pi^+)\ ,\nonumber\\
&& \Gamma[\Xi_c(2970)^+\to \Xi_c(2645)^+\pi^0] = \frac{(g_{A,1}^{+H})^2}{12\pi f_0^2}F_1(M_i,E_f;\pi^0) \  ,  \nonumber\\
\end{eqnarray}
for Fig.~\ref{fig:XiDecay} (c). In those equations, the notation is understood similarly to that used in Eq.~(\ref{DecayFormula2}). From the formulae we can find that, for instance, the decay width of $\Xi_c(2970)^0\to \Sigma_c(2455)^0\bar{K}^0$ is twice that of $\Xi_c(2970)^0\to \Sigma_c(2455)^+{K}^-$. Such a difference is understood by the Clebsch-Gordan coefficients with respect to the isospin. In addition, similarly to the formulae for $\Lambda_c(2765)$ demonstrated in Sec.~\ref{sec:Analysis}, the $SU(2)_h$ heavy quark spin symmetry is manifestly seen.

Similarly, the analytic formulae for decay widths of $\Xi_c(2645)$ corresponding to the diagram in Fig.~\ref{fig:XiC1Decay} are
\begin{eqnarray}
&& \Gamma[\Xi_c(2645)^0\to\Xi^+_c\pi^-] =\frac{(g_{A,1}^{+L})^2}{12\pi f_0^2}F_2(M_i,E_f;\pi^-)  \ ,\nonumber\\
&& \Gamma[\Xi_c(2645)^0\to\Xi^0_c\pi^0] = \frac{(g_{A,1}^{+L})^2}{24\pi f_0^2}F_2(M_i,E_f;\pi^0)  \ ,\nonumber\\
&& \Gamma[\Xi_c(2645)^+\to\Xi^0_c\pi^+] = \frac{(g_{A,1}^{+L})^2}{12\pi f_0^2}F_2(M_i,E_f;\pi^+) \  ,  \nonumber\\
&& \Gamma[\Xi_c(2645)^+\to\Xi^+_c\pi^0] = \frac{(g_{A,1}^{+L})^2}{24\pi f_0^2}F_2(M_i,E_f;\pi^0) \,  ,  
\end{eqnarray}
where the notation is the same as in Eq.~(\ref{SigmaDecay2}). Again differences by a factor of two appear due to the Clebsch-Gordan coefficients with respect to the isospin.

\bibliography{reference}

\end{document}